\documentclass[prd,twocolumn,tightenlines,superscriptaddress,nofootinbib]{revtex4}
\usepackage{amssymb,latexsym}
\usepackage{amsmath,amsbsy,bbm}
\usepackage{epsfig,bm}
\usepackage{graphicx,comment}
\usepackage{color}
\usepackage[normalem]{ulem}
\begin{document}

\title{Solving the puzzle of an unconventional phase transition for a 2d dimerized quantum Heisenberg model}

\author{F.-J. Jiang}
\email[]{fjjiang@ntnu.edu.tw}
\affiliation{Department of Physics, National Taiwan Normal University, 
88, Sec.4, Ting-Chou Rd., Taipei 116, Taiwan}

\vspace{-2cm}
  
\begin{abstract}
Motivated by the indication of a new critical theory for the
spin-1/2 Heisenberg model with a spatially staggered anisotropy on the
square lattice as suggested in \cite{Wenzel08}, we re-investigate
the phase transition of this model induced by dimerization using first principle Monte
Carlo simulations. We focus on studying the finite-size scaling of
$\rho_{s1} 2L$ and $\rho_{s2} 2L$, where $L$ stands for the spatial box size used in the 
simulations and $\rho_{si}$ with $i \in \{1,2\}$ is the spin-stiffness in the $i$-direction. 
Remarkably, while we do observe a large correction to scaling for the observable
$\rho_{s1}2L$ as proposed in \cite{Fritz11}, the data for $\rho_{s2}2L$ exhibit a good scaling behavior
without any indication of a large correction. 
As a consequence, we are able to obtain a numerical value for the critical exponent 
$\nu$ which is consistent with the known $O(3)$ result with moderate computational effort. 
Specifically, the numerical 
value of $\nu$ we determine by fitting the data points of $\rho_{s2}2L$ to their expected 
scaling form is given by $\nu=0.7120(16)$, which agrees quantitatively with the most 
accurate known Monte Carlo $O(3)$ result $\nu = 0.7112(5)$. Finally, while we can 
also obtain a result of $\nu$ from the observable second Binder
ratio $Q_2$ which is consistent with $\nu=0.7112(5)$, the uncertainty of $\nu$ calculated from $Q_2$ is more than twice
as large as that of $\nu$ determined from $\rho_{s2}2L$.

\end{abstract}

\maketitle

{\bf Introduction}.---
Heisenberg-type models have been one of the central research topics in condensed 
matter physics during the last two decades. The reason that these models 
have triggered great theoretical interest is twofold.
First of all, Heisenberg-type models are relevant to real materials.
Specifically, the spin-1/2 Heisenberg model on the square 
lattice is the appropriate model for understanding the undoped precursors of 
high $T_c$ cuprates (undoped antiferromagnets).
Second, because of the availability of efficient Monte Carlo algorithms as well
as the increased power of computing resources, properties of undoped antiferromagnets
on geometrically non-frustrated lattices can be simulated with unprecedented 
accuracy \cite{Beard96,Sandvik97,Sandvik99,Kim00,Wang05,Jiang08,Wenzel09}. As a 
consequence, these models are particular suitable for examining theoretical predictions 
and exploring ideas. For instance, Heisenberg-type models
are often used to examine field theory predictions regarding the universality class of
a second order phase transition \cite{Wang05,Wenzel08,Wenzel09}. Furthermore, a new proposal of determining the low-energy constant, 
namely the spinwave velocity
$c$ of antiferromagnets with $O(2)$ and $O(3)$ symmetry, through the squares of temporal and spatial 
winding numbers was verified to be valid and this idea has greatly improved the accuracy
of the related low-energy constants \cite{Jiang11.1,Jiang11.2}.
On the one hand, Heisenberg-type models on geometrically non-frustrated lattices are
among the best quantitatively understood condensed matter physics systems; on the other
hand, despite being well studied, several recent numerical investigation of spatially anisotropic 
Heisenberg models have led to unexpected results \cite{Wenzel08,Pardini08,Jiang09.1}. 
In particular, Monte Carlo evidence indicates that the anisotropic
Heisenberg model with staggered arrangement of the antiferromagnetic 
couplings may belong to a new universality class, in contradiction
to the theoretical $O(3)$ universality prediction \cite{Wenzel08}.
For example, while the most accurate Monte Carlo value for the critical exponent
$\nu$ in the $O(3)$ universality class is given by $\nu=0.7112(5)$ \cite{Cam02},
the corresponding $\nu$ determined in \cite{Wenzel08} is $\nu=0.689(5)$. 
Although the subtlety of calculating the critical exponent $\nu$ from performing
finite-size scaling analysis has been demonstrated for a similar anisotropic 
Heisenberg model on the honeycomb lattice \cite{Jiang09.2}, the discrepancy between $\nu = 0.689(5)$ 
and $\nu=0.7112(5)$ observed in \cite{Wenzel08,Cam02} remains to be understood.

In order to clarify this issue further, several efforts have been devoted to studying
the phase transition of this model induced by dimerization. For instance, an 
unconventional finite-size scaling is proposed in \cite{Jiang10.1}. Further, it is 
argued that there is a large correction to scaling for this phase 
transition which leads to the unexpected $\nu = 0.689(5)$ obtained in \cite{Wenzel08}. 
Still, direct numerical evidence to solve this puzzle is not
available yet. In this study, we undertake the challenge of determining the critical exponent $\nu$
by simulating the spin-1/2 Heisenberg model with
a spatially staggered anisotropy on the square lattice. 
The relevant observables considered in this study for calculating the 
critical exponent $\nu$ are $\rho_{s1}2L$, $\rho_{s2}2L$ and $Q_2$. Here
$\rho_{si}$ with $i \in \{1,2\}$ are the spin stiffness in the $i$-direction,
$L$ is the box size used in the simulations and $Q_2$ is the second Binder ratio
which will be defined later. 
Further, we analyze in more detail
the finite-size scaling of $\rho_{s1} 2L$ and $\rho_{s2} 2L$.
The reason that $\rho_{s1} 2L$ and $\rho_{s2} 2L$ are chosen is twofold. 
First of all, these two observables can be calculated
to a very high accuracy using loop algorithms \cite{Beard96}. Second, one can measure 
$\rho_{s1}$ and $\rho_{s2}$ separately. 
On isotropic systems, one would
naturally use $\rho_s$, which is the average of $\rho_{s1}$ and $\rho_{s2}$,
for the data analysis.
However, for the anisotropic model considered here, 
we find it is useful to analyze both the data of $\rho_{s1}$ and $\rho_{s2}$
because such a study may reveal the impact of anisotropy on the system.
Surprisingly, as we will show later, the observable $\rho_{s2} 2L$ receives a much
less severe correction than $\rho_{s1} 2L$.
Hence $\rho_{s2} 2L$ is a better suited quantity than
$\rho_{s1} 2L$ for the finite-size scaling analysis. Indeed, with moderate 
computational effort, 
we can obtain a numerical value for $\nu$ consistent with the most accurate 
$O(3)$ Monte Carlo result $\nu = 0.7112(5)$.
 
This paper is organized as follows. First, the anisotropic
Heisenberg model and the relevant observables studied in this work are briefly described after which we 
present our numerical results.
In particular, the corresponding critical point as well as the critical
exponent $\nu$ are determined by fitting the numerical data to their predicted
critical behavior near the transition. A final section then concludes 
our study.

\begin{figure}
\begin{center}
\includegraphics[width=0.30\textwidth]{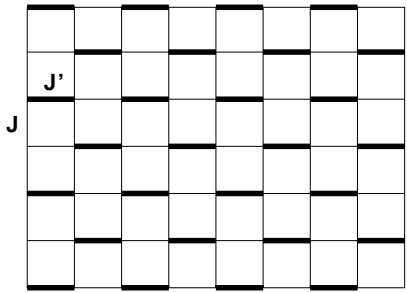}
\end{center}\vskip-0.5cm
\caption{The spatially anisotropic Heisenberg model considered in this study.}
\label{fig1}
\end{figure}

{\bf Microscopic Model and Corresponding Observables}.---
The Heisenberg
model considered in this study is defined by the Hamilton operator
\begin{eqnarray}
\label{hamilton}
H = \sum_{\langle xy \rangle}J\,\vec S_x \cdot \vec S_{y}
+\sum_{\langle x'y' \rangle}J'\,\vec S_{x'} \cdot \vec S_{y'},
\end{eqnarray}
where $J'$ and $J$ are antiferromagnetic exchange couplings connecting
nearest neighbor spins $\langle  xy \rangle$
and $\langle x'y' \rangle$, respectively. Figure 1 illustrates the Heisenberg
model described by Eq.~(\ref{hamilton}). 
To study the critical behavior of this anisotropic Heisenberg model near 
the transition driven by the anisotropy, in particular, to determine 
the critical point as well as the critical exponent $\nu$, 
the spin stiffnesses in the $1$- and $2$-directions which are defined by\vskip-0.5cm
\begin{eqnarray}
\rho_{si} = \frac{1}{\beta L^2}\langle W^2_{i}\rangle,
\end{eqnarray}
are measured in our simulations.
Here $\beta$ is the inverse temperature and $L$ again 
refers to the spatial box size. Further $\langle W^2_{i} \rangle$ 
with $i \in \{1,2\}$ is
the winding number squared in the $i$-direction.
In addition, the second Binder ratio $Q_2$, which is defined by
\begin{equation}
Q_2 = \frac{\langle (m_s^z)^2\rangle^2}{\langle (m_s^z)^4\rangle}
,\end{equation}
is also measured in our simulations as well. Here $m_s^z$ is 
the $z$-component of the staggered magnetization 
$\vec{m}_s = \frac{1}{L^2}\sum_{x}(-1)^{x_1 + x_2}\vec{S}_x$.
By carefully investigating the spatial volume 
and the $J'/J$ dependence of
$\rho_{s i}L$ as well as $Q_2$, one can determine the critical point and the critical exponent $\nu$ with high precision.

\begin{table}[ht]
\label{tab1}
\begin{center}
\begin{tabular}{ccccc}
\hline
{\text{observable}}& $\,L\, $  & $\nu$ & $(J'/J)_c$ & $\chi^2/{\text{DOF}}$\\
\hline
\hline
$\rho_{s2}2L$  & $48\, \le \,L\, \le\, 96$  & 0.7150(28) & 2.51951(8) & 1.0\\
\hline
$\rho_{s2}2L^{\star}$  & $48\, \le \,L\, \le\, 96$  & 0.7095(32) & 2.51950(8) & 0.9\\
\hline
$\rho_{s2}2L$  & $48\, \le \,L\, \le\, 136$  & 0.7120(16) & 2.51950(3) & 1.1\\
\hline
$\rho_{s2}2L$  & $60\, \le \,L\, \le\, 136$  & 0.7120(18) & 2.51950(3) & 1.1\\
\hline
$\rho_{s2}2L$  & $66\, \le \,L\, \le\, 136$  & 0.7125(20) & 2.51950(4) & 1.1\\
\hline
$\rho_{s2}2L^{\star}$  & $48\, \le \,L\, \le\, 136$  & 0.7085(16) & 2.51950(3) & 0.9\\
\hline
$\rho_{s2}2L^{\star}$  & $60\, \le \,L\, \le\, 136$  & 0.7087(17) & 2.51950(3) & 0.9\\
\hline
$\rho_{s2}2L^{\star}$  & $66\, \le \,L\, \le\, 136$  & 0.7096(19) & 2.51950(4) & 0.9\\
\hline
$\rho_{s1}2L$  & $24\, \le \,L\, \le\, 80$ & 0.689(3)  & 2.5194(4) & 1.2\\
\hline
$\rho_{s1}2L^{\star}$  &$24\, \le \,L\, \le\, 80$ & 0.683(4)  & 2.5194(3) & 1.1\\
\hline
$\rho_{s1}2L$ &   $48\, \le \,L\, \le\, 136$  & 0.701(3) & 2.5194(3) & 1.6\\
\hline
$Q_2\,$&  $48\, \le \,L\, \le\, 136$  & 0.7116(50) & 2.51952(8) & 1.4\\
\hline
$Q_2^{\star}\,$&  $48\, \le \,L\, \le\, 136$  & 0.7050(48) & 2.51950(8) & 1.3\\
\hline
\hline
\end{tabular}
\end{center}
\caption{The numerical values for $\nu$ and $(J'/J)_c$ calculated from $\rho_{s2}2L$, $\rho_{s1}2L$ 
and $Q_2$. All results are obtained using a second order Taylor series expansion of Eq.~(\ref{FSS}), except those
with a star, which are determined using an expansion to third order. The confluent correction $\omega$ 
is included in the fit explicitly only for $\rho_{s1}2L$.
}
\end{table}

\begin{figure}
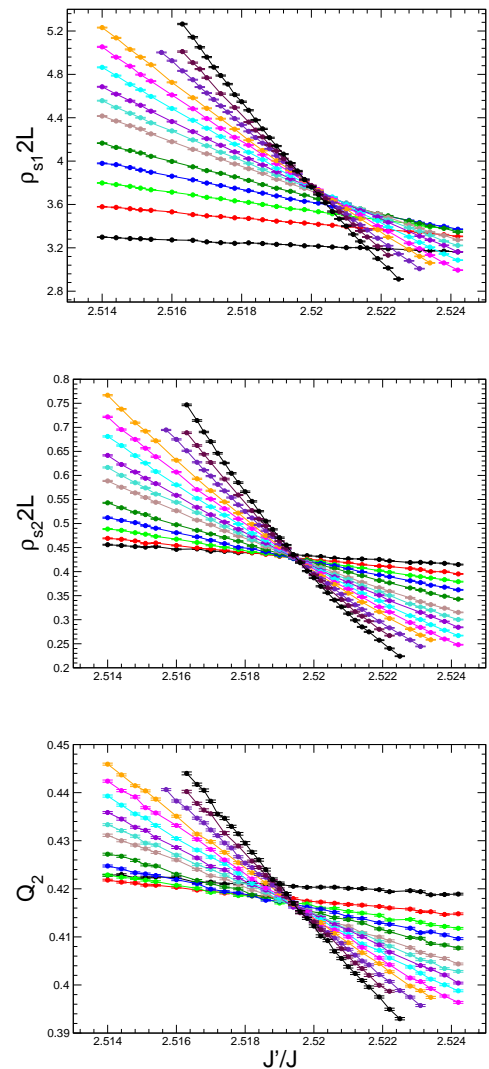

\begin{center}
\vbox{
\includegraphics[width=0.35\textwidth]{rhos12L.eps}
\vskip0.45cm
\includegraphics[width=0.35\textwidth]{rhos22L.eps}
\vskip0.45cm
\includegraphics[width=0.35\textwidth]{2nd_binder_ratio.eps}
}
\end{center}
\caption{Monte Carlo data of $\rho_{s1}2L$ (top), $\rho_{s2}2L$ 
(middle), and $Q_2$ (bottom).}
\label{fig2}
\end{figure}

\begin{figure}
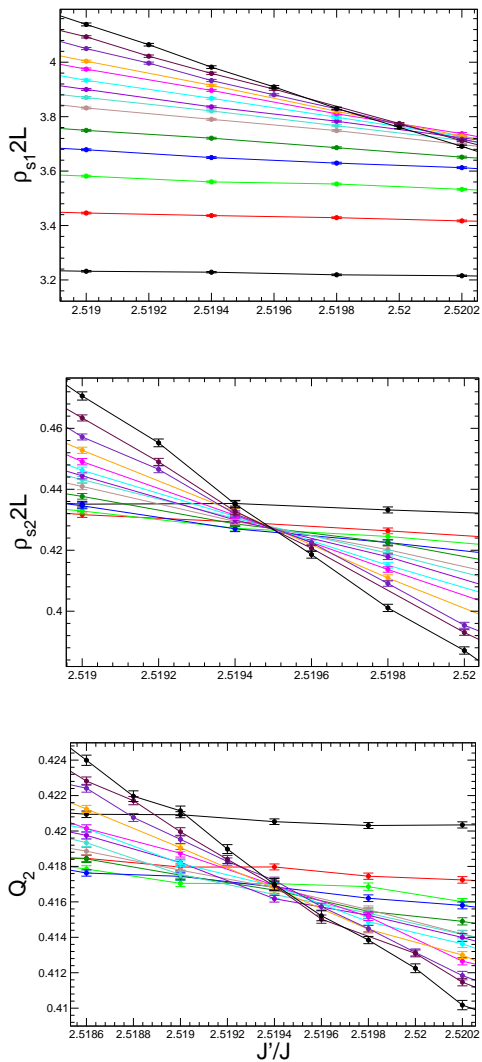

\begin{center}
\vbox{
\includegraphics[width=0.35\textwidth]{crossing_rhos12L.eps}
\vskip0.5cm
\includegraphics[width=0.35\textwidth]{crossing_rhos22L.eps}
\vskip0.5cm
\includegraphics[width=0.35\textwidth]{crossing_2nd_binder_ratio.eps}
}
\end{center}
\caption{Crossing of $\rho_{s1}2L$ (top), $\rho_{s2}2L$ (middle), and $Q_2$ (bottom) 
for different $L$ near the transition.}
\label{fig3}
\end{figure}

{\bf Determination of the Critical Point and the Critical Exponent $\nu$}.---
To study the quantum phase transition,
we have carried out large scale Monte Carlo simulations using a loop algorithm.
Further, to calculate the relevant critical exponent $\nu$ and to determine the
location of the critical point in the parameter space $J'/J$, we have employed 
the technique of finite-size scaling for certain observables. For example,
if the transition is second order, then near the transition the observable 
$\rho_{si} 2L$ for $i\in \{1,2\}$ and $Q_2$ should be described well by the following finite-size scaling ansatz
\begin{equation}
\label{FSS}
{\cal O}_{L}(t) = ( 1 + bL^{-\omega} )g_{{\cal O}}(tL^{1/\nu},\beta/L^z), 
\end{equation}
where ${\cal O}_{L}$ stands for $\rho_{si}2L$ with $i \in \{1,2\}$ or $Q_2$, 
$t = (j_c-j)/j_c$ with $j = (J'/J)$, $b$ is some constant,
$\nu$ is the critical exponent corresponding to the correlation length $\xi$, 
$\omega$ is the confluent correction exponent and $z$ is the dynamical 
critical exponent which is $1$ for the transition 
considered here. Finally, $g_{{\cal O}}$ is a 
smooth function of the variables $tL^{1/\nu}$ and $\beta/L^z$. From Eq.~(\ref{FSS}), 
one concludes that the curves for ${\cal O}_{L}$ corresponding to different $L$, as functions of $J'/J$, should intersect 
at the critical point $(J'/J)_c$ for large $L$. 
To calculate the critical exponent $\nu$ and the critical point $(J'/J)_c$,
in the following we will apply the finite-size scaling formula,
Eq.~(\ref{FSS}), to $\rho_{s1} 2L$, $\rho_{s2} 2L$ as well as $Q_2$. 
Without loss of generality, we have 
fixed $J=1$ in our simulations and have varied $J'$. Additionally, the box size used in 
the simulations ranges from $L = 16$ to $L = 136$.
To reach a lattice size as large as possible,
we use $\beta J = 2L$ for each $L$ in our simulation. As a result, the temperature dependence
in Eq.~(\ref{FSS}) drops out. Figure \ref{fig2} shows the Monte Carlo data for 
$\rho_{s1} 2L$, $\rho_{s2} 2L$ and $Q_2$ as functions of $J'/J$. 
The figure clearly indicates that the phase 
transition is most likely second order since for all the observables $\rho_{s1} 2L$, $\rho_{s2} 2L$ and $Q_2$, the curves of different $L$
tend to intersect near a particular point in 
the parameter space $J'/J$. The most striking observation from our results
is that the observable $\rho_{s1} 2L$ receives a much more severe correction 
than $\rho_{s2} 2L$. This can be understood from the trend of the crossing among these 
curves for different $L$ near the transition (figure \ref{fig3}). Therefore one expects that a better 
determination of $\nu$ can be obtained by applying the finite-size scaling ansatz 
Eq.~(\ref{FSS}) to $\rho_{s2} 2L$. Before presenting our results,
we would like to point out 
that data from large volumes are essential 
in order to determine the critical exponent $\nu$ accurately
as was emphasized in \cite{Jiang09.2}. We will use the strategy employed
in \cite{Jiang09.2} for our data analysis as well. 
Let us first focus on $\rho_{s2}2L$ since this observable shows a
good scaling behavior. Notice from figure \ref{fig3}, the trend of crossing for different $L$ of $\rho_{s2}2L$ indicates
that the confluent correction is negligible for lattices of larger size. 
Therefore one expects that a result consistent with the 
theoretical prediction can be reached with $b = 0$ in formula~(\ref{FSS}) if data
from large $L$ are taken into account in the fit. Indeed with a Taylor expansion of
Eq.~(\ref{FSS}) up to second order in $tL^{1/\nu}$ as well as letting $b = 0$ 
in Eq.~(\ref{FSS}), we arrive at $\nu = 0.7120(16)$ and $(J'/J)_c = 2.51950(3)$ using
the data of $\rho_{s2} 2L $ with $48 \le L \le 136$. In obtaining the results $\nu = 0.7120(16)$ and 
$(J'/J)_c = 2.51950(3)$, we have performed bootstrap sampling on the raw data and have carried out
a large number (around 1000) of fits. 
The inclusion of higher order terms in the Taylor series and eliminating data of smaller $L$
in the fits leads to compatible (and consistent) results with what we have just obtained.
Notice that the $\nu$ we obtain is consistent with the most
accurate Monte Carlo result $\nu=0.7112(5)$ in the $O(3)$ universality class. Further,
the critical point $(J'/J)_c = 2.51950(3)$ we calculate agrees with the known results in
the literature \cite{Wenzel08,Jiang10.1} as well.

Having determined $\nu=0.7120(16)$ using the observable $\rho_{s2}2L$, we turn to the 
calculations of $\nu$ based on the observables $\rho_{s1}2L$ and $Q_2$. First of all, we would like to reproduce the 
unexpected result 
$\nu=0.689(5)$ found in \cite{Wenzel08}. Indeed, using the Monte Carlo data of $\rho_{s1}2L$
with $L$ ranging from $L=24$ to $L=80$ (the size of $L=80$ is similar to the 
largest lattice 
($L=72$) used in \cite{Wenzel08} in obtaining $\nu=0.689(5)$), we arrive at $\nu=0.689(3)$ and
$(J'/J)_c=2.5194(4)$, both of which are statistically consistent with 
those determined in \cite{Wenzel08}. Further, a numerical value for $\nu$ 
consistent with its theoretical prediction $\nu=0.7112(5)$
could never have been obtained using the available data for $\rho_{s1}2L$ . 
This implies that the correction to scaling for $\rho_{s1}2L$ is large 
and the lattice data for $\rho_{s1}2L$ with larger $L$ are required to reach a numerical 
value of $\nu$ consistent with the theoretical expectation. 
Finally, for the observable $Q_2$, using the leading finite-size scaling ansatz (i.e. letting $b = 0$ in Eq.~(\ref{FSS})), 
we are able to reach a value for $\nu$ which agrees even quantitatively with $\nu=0.7112(5)$. However,
the uncertainty of $\nu$ calculated from $Q_2$ is more than twice as large as that of $\nu$ determined from $\rho_{s2}2L$. 
\vskip0.07cm
The results for $\nu$ and $(J'/J)_c$ calculated from our finite-size scaling analysis are summarized in table 1.

\vskip0.07cm
{\bf Conclusions}.--- 
In this paper, we re-investigated the critical behavior at the phase transition 
induced by dimerization of the spin-1/2 Heisenberg model with a spatially 
staggered anisotropy. Unlike the scenario suggested in \cite{Wenzel08} that 
an unconventional universality class is observed, we 
conclude that indeed this second order phase transition is well described 
by the $O(3)$ universality class. Our observation of $\rho_{s2}2L$ being a 
good observable for determining the critical exponent $\nu$ is crucial for
reading this conclusion by confirming the $O(3)$ critical exponent for this
phase transition with high precision. While we do observed a large correction to scaling
for the observable $\rho_{s1}2L$ as proposed in \cite{Fritz11}, 
the data points of 
$\rho_{s2}2L$ show good scaling behavior. Specifically, with $\rho_{s2}2L$, we can easily 
reach a highly accurate numerical value for $\nu$ consistent with the theoretical 
predictions without taking the confluent correction into account in the fit.
The large correction to scaling observed for $\rho_{s1}2L$ in principle should
influence all observables. Hence the most reasonable explanation for the good scaling
behavior of $\rho_{s2}2L$ shown here is that the prefactor $b$ in Eq.~(\ref{FSS}) 
for $\rho_{s2}2L$ is very small. As a result, we are able to determine the expected
numerical value for $\nu$ using data of $\rho_{s2}2L$ with moderate lattice sizes.
Still, a more rigorous theoretical study such as investigating whether there exists
a symmetry that protects $\rho_{s2}2L$ from being affected by the large correction 
to scaling as suggested in \cite{Fritz11} will be an interesting topic to 
explore. For example, in \cite{Fritz11} it is argued that 
the enhanced correction to scaling observed for this phase transition
might be due to a cubic irrelevant term which contains one-derivative in the 
1-direction. 
The first thing one would like to understand is whether
the feature of this irrelevant term, namely it contains one-derivative
in the 1-direction, will lead to our observation that the large 
correction to scaling has little impact on $\rho_{s2}2L$.
In summary, here we 
present convincing numerical evidence to support that the phase transition
considered in this study is well described by the $O(3)$ 
universality class prediction, at least for the critical exponent $\nu$
which is investigated in detail in this study. 
Finally, whether the good scaling of $\rho_{s2}2L$ observed here is a coincidence 
or is generally applicable for quantum Heisenberg models with a similar spatially anisotropic 
pattern, remains an interesting topic for further investigation.
\vskip0.07cm
{\bf Acknowledgements}.---
We thank B. Smigielski for correcting the manuscript for us.
Useful discussions with S. Wessel, M. Vojta, and U.-J. Wiese is acknowledged. 
Part of the simulations 
in this study were based on the loop algorithms available in ALPS 
library \cite{Troyer08}. This work is partially supported by 
NSC (Grant No. NSC 99-2112-M003-015-MY3) and NCTS (North) of R. O. C..

\end{document}